\documentclass[prl,preprint,showpacs,superscriptaddress]{revtex4}
\usepackage{graphicx}
\usepackage{dcolumn}

\def\Qtens{\mbox{\sffamily\bfseries Q}}

\def\Itens{\mbox{\sffamily\bfseries I}}

\def\ftens{\mbox{\sffamily\bfseries f}}
\def\vtens{\mbox{\sffamily\bfseries v}}
\def\nabvec{\mbox{\boldmath $\nabla$}}

\def\beq{\begin{equation}}
\def\bea{\begin{eqnarray}}
\def\eeq{\end{equation}}
\def\eea{\end{eqnarray}}

\begin{document}

\title{Active nematics on a substrate: giant
number fluctuations and long-time tails}

\author{Sriram Ramaswamy}
\email{sriram@physics.iisc.ernet.in}
\author{R. Aditi Simha}
\email{aditi@physics.iisc.ernet.in}
\affiliation{Centre for Condensed-Matter Theory, Department of
Physics, Indian Institute of Science, Bangalore 560 012, India}
\author{John Toner}
\email{jjt@darkwing.uoregon.edu}
\affiliation{Department of Physics, University of Oregon, Eugene OR
94703 USA}

\date{version of 6 June 2002, printed \today}

\begin{abstract}

We construct the equations of motion for the coupled dynamics of
order parameter and concentration for the nematic phase of driven
particles on a solid surface, and show that they imply (i) giant
number fluctuations, with a standard deviation proportional to the
mean and (ii) long-time tails $\sim t^{-d/2}$ in the autocorrelation of
the particle velocities in $d$ dimensions despite the absence of a
hydrodynamic velocity field. Our predictions can be tested in
experiments on aggregates of amoeboid cells as well as on layers of
agitated granular matter. \end{abstract}

\pacs{ }
\maketitle

Recent experiments \cite{gruler,kudro} offer evidence for long-ranged
orientational order in collections of driven particles. The work of
\cite{gruler} is concerned with aggregates of living cells confined to a
solid surface, whereas that of \cite{kudro} studies vertically vibrated
layers of granular rods. Two kinds of uniaxial orientational order are
discussed in \cite{gruler}, and appear to arise in \cite{kudro} as well:
(i) {\em nematic} or {\em apolar}, where the axes of the elongated cells
are oriented, say along a mean direction $\hat{\bf n}$, but
$\hat{\bf n}$ and - $\hat{\bf n}$ are equivalent; and (ii) {\em
vectorial} or {\em polar} order, where
$\hat{\bf n}$ and - $\hat{\bf n}$ are not equivalent. On symmetry
grounds \cite{curieprincip}, a vector-ordered 
nonequilibrium steady state should have a nonzero macroscopic 
drift velocity ${\bf v}_0$, while for true nematic 
order ${\bf v}_0 = 0$.

We shall be concerned here with strictly nematic order. In \cite{kudro}
a nematic phase is mentioned but not discussed much. In \cite{gruler},
there is particularly convincing evidence for a nematic, including images 
of classic strength -$1/2$ disclinations, from studies on melanocytes
(cells which distribute pigment in the skin). In fact in this class
of cells the individual particles themselves are head-tail symmetric, or
{\em apolar}, so that the drift velocity is zero for each particle. The
cells involved are tens to hundreds of microns in length, and the rods
in \cite{kudro} are 6 mm long, so that thermal Brownian motion plays
no role in the dynamics of these particles. What distinguishes the
particles in these experiments from conventional nematics is their
{\em active} or {\em driven} character: they absorb energy from their
surroundings (from nutrient in one case, shaking in the other) and
transform it into motion in the plane. The dynamical behaviour of
melanocytes, for example, involves rhythmic movements of the
cell body as well as two long projections called dendrites 
which appear to be responsible for the inter-cell
interaction and the level of fluctuations \cite{gruler}. In granular
rods the interaction and noise come from collisions. Ordered
phases of particles on an air-table \cite{ojha} would also obey our
model, except for possible complications arising from the periodic
array of air-holes.

As has been noted earlier
\cite{vicseketal,tonertu} the organization, dynamics, and fluctuation
properties of such active, energy-dissipating objects cannot be
described by equilibrium statistical physics. Continuum models
appropriate for the dynamics of {\em polar} order wherein the
particles are on average in motion were presented in \cite{tonertu}; the
extension of those models to include momentum-conserving
hydrodynamic flow and both polar and apolar order was discussed in
\cite{sppprl}. The cells in \cite{gruler} and the rods in \cite{kudro},
however, move on a solid surface which acts as a momentum sink for
the velocity even at zero wavenumber.

In this paper we derive the most general {\it universal} equations of
motion, valid for all systems of active nematogenic particles {\it
without} total momentum conservation
\cite{momentum}, for the {\it apolar} order parameter and
concentration fields in such systems,  
retaining all relevant terms allowed by symmetry. Our aim is
to look for {\it universal} properties, independent of microscopic
details, that distinguish these active nematics from their counterparts
at thermal equilibrium.

Two predictions stand out as most striking: (i) The standard deviation
in the number $N$ of particles is {\it enormous,} scaling as $N$ in the
entire nematic phase for two dimensional systems.   By contrast, in {\it
any} thermal equilibrium system with a finite compressibility, which in
practice means any equilibrium system not at a continuous phase
transition, number fluctuations are far smaller, scaling as $\sqrt{N}$ as
$N \rightarrow \infty$. (ii) The autocorrelation of the velocity of a
tagged particle decays with time $t$ as $t^{-d/2}$ for space dimension
$d \geq 2$. It should be recalled that precisely this form of long-time
tail \cite{tails} occurs in bulk fluids at thermal equilibrium, but as a
result of advection of particles by thermal fluctuations in the
hydrodynamic velocity field, which decays slowly because of
momentum conservation. Here the total momentum of the particles is
not conserved, since it is damped by friction with the substrate
\cite{srmaz}. A collection of particles at thermal equilibrium on a
substrate would show velocity autocorrelations decaying much more
rapidly \cite{pago}; specifically as $t^{-(d + 2)/2}$. It should be
possible to test these predictions by studying large-scale real-space
images and videos of the behavior of systems such as those in
\cite{gruler} as well as two-dimensional nematic phases in agitated
layers of granular particles \cite{kudro}.

Our predictions for {\em nematically ordered} phases of powders stand in
dramatic contrast to experimental results obtained on {\em disordered}, 
monodisperse powders (ping-pong balls) 
\cite{ojha}, which showed that the latter systems could be described
surprisingly completely by equilibrium statistical mechanics, albeit
with a very high fictive temperature \cite{liudurian}. The findings we've summarized
above clearly show that nematic phases of rod-like powders {\it
cannot} be described by equilibrium statistical mechanics {\it at all},
even at the crudest level of scaling \cite{liukurchan}.

We turn next to the construction of our equations of motion.

Since our nematogens are spread on a solid substrate, there is only one
conserved quantity of relevance, namely, the number of particles. A
description valid at sufficiently long length- and time-scales \cite{mpp}
needs then to include only the concentration field $c({\bf r},t)$ of
particles and the orientational order parameter, described by the
traceless symmetric tensor field $\Qtens$ with components
$Q_{ij}({\bf r},t)$ at point
${\bf r}$ and time $t$. For uniaxial nematics, to which we restrict our
attention here, we can write
$\Qtens = [{\bf n} {\bf n} - (1/d)\Itens]S$, where the unit vector ${\bf
n}$ is the director field, $\Itens$ is the unit tensor, and the conventional 
scalar order parameter $S$
\cite{degp} measures the degree of order. It is however convenient to
introduce and later eliminate a fast variable, the velocity field ${\bf
v}({\bf r},t)$ of the particles. Number conservation tells us that \beq
\label{contin} {\partial c \over \partial t} = - \nabla \cdot {\bf j} \eeq
where the number current ${\bf j} = c {\bf v}$. Newton's 2nd law for
the local momentum density $m {\bf j}$, where $m$ is the mass of a
particle, gives
\beq
\label{momentum} m{\partial {\bf j} \over \partial t} = -\Gamma {\bf
v} - {\bf \nabla} \cdot {\bf \sigma}+{\bf f}_R = -\Gamma \ \vtens -
w_o \nabla c - w_1 \nabla \cdot (\Qtens c) + \ftens_R \eeq up to
bilinear order in the fields. In (\ref{momentum}), the term in
$\Gamma$ represents friction with the substrate, the stress tensor
$\sigma$ contains the effect of interparticle interactions, the $w_0$
term can be thought of as osmotic pressure, ${\bf f}_R$ is a random,
nonconserving, spatiotemporally white Gaussian noise, not in general
of thermal origin, and the $w_1$ term, unique to this driven system,
says that inhomogeneities in the nematic order drive mass motion.

To see where this comes from, note first that, by definition,
inhomogeneities in ${\bf \sigma}$ produce local acceleration. What is
novel here is that there is a contribution to the stress tensor
proportional to the nematic order parameter. This is allowed by
symmetry here, and can be obtained by noting (see
\cite{sppprl,pumplong}) that each active particle is a force dipole. Such
a term is {\it not} allowed in an {\it equilibrium} nematic since the
stress tensor in equilibrium must be derived from functional
derivatives of the free energy. Since {\it that } free energy must be
rotation invariant, it can only depend on {\it derivatives} of the
nematic director ${\bf \hat{n}}$ (in fact, this argument leads to the
familiar Frank free energy \cite{degp} for equilibrium nematics).
Functional derivatives of this then lead to a stress ${\bf \sigma}$
which depends on {\it derivatives} of {\bf n}. Since we are considering
{\it non-equilibrium} nematics, however, this constraint that ${\bf
\sigma}$ be derived from a functional derivative of a
rotation-invariant free energy does not apply. Hence, the only
constraint on allowed terms is rotation invariance of the equations of
motion, which is respected by a term in ${\bf \sigma}$ proportional to
{\bf Q} \cite{high}, since both objects are symmetric second rank
tensors.

We now restrict our attention only to small fluctuations about the
uniaxial nematic phase, aligned along the
$z$ axis, and denote vector components in directions normal to $z$ by
the subscript $\perp$. We assume the nematic order is well developed,
so that
$S$ can be treated as constant, and consider spatial variations $c({\bf
r},t) = c_0 + \delta c({\bf r},t)$, where $c_0$ is the mean
concentration, and
${\bf n({\bf r},t)} = \hat{\bf z} + {\bf \delta n({\bf r},t)}$. Since ${\bf
n}$ is a unit vector, we need consider only the transverse components
${\bf \delta n}_{\perp}$.

To close our equations, we need the equation of motion for {\bf n}
\cite{sppprl}, which is identical in form to that for an equilibrium
nematic and reads, to linear order:
\begin{eqnarray}
\partial_t\delta {\bf n}_{\perp} = \lambda_{+}
\partial_z {\bf v}_{\perp} + \lambda_{-}
\nabvec_{\perp} v_z + K_1 \nabvec_{\perp}(\nabvec_{\perp}\cdot \delta
{\bf n}_{\perp}) + K_2 \nabla^2_{\perp}\delta {\bf n}_{\perp} + K_3
\delta^2_z\delta {\bf n}_{\perp}+ {\bf f}_{\perp}. \label{neom1}
\end{eqnarray} Here $\lambda_{\pm}= (\lambda \pm 1)/2\,\,$;
$\,\lambda$ is the ``flow alignment parameter'', familiar from 
equilibrium nematics, which affects the response of the nematic
director to shear, and the noise ${\bf f}_{\perp}$ is delta-correlated in
space and time: \begin{eqnarray}
\left< f_{\perp i} ({\bf r}, t) f_{\perp j} 
({\bf r^{\prime}}, t^{\prime})\right> =
\delta_{ij}\,\Delta_n \delta^d({\bf r- r^{\prime}}) \delta (t -t^{\prime})
\label{noise}
\end{eqnarray} We now use the equation of motion (2) to eliminate
{\bf v}. For hydrodynamically slow processes, $\partial_t {\bf v} \ll
\Gamma {\bf v}$, and so we can neglect the $\partial_t {\bf v}$ term
in (\ref{momentum}) and immediately solve for {\bf v}, finding, to
linear order,
\begin{eqnarray} {\bf v} = -\alpha \left(\,\partial_z \delta {\bf
n}_{\perp} + \left(\nabla_{\perp} \cdot \delta {\bf
n}_{\perp}\right) {\bf \hat{z}}\right) - \gamma_1 \nabvec_{\perp} c 
- \gamma_2\, \partial_z\, c\,{\bf \hat{z}}+{{\bf f}_R \over
\Gamma} \label{vsol}
\end{eqnarray} where $\alpha$ and $\gamma_i$ are related to 
$c_o$, $S$ and the coefficients in (\ref{momentum}).  
Inserting (\ref{vsol})
into (\ref{contin}) gives, to linear order in fluctuations, the
concentration equation of motion \beq
\label{concsmallfluc} {\partial \delta c \over \partial t} = (D_z
\partial_z^2 + D_{\perp} \nabla_{\perp}^2)\,\delta c + 2 c_0\,\alpha
\,\partial_z (\nabla_{\perp} \cdot {\bf \delta n}_{\perp}) + \nabla \cdot
{\bf f}_c
\eeq where the various coefficients in (\ref{concsmallfluc}) and the
noise ${\bf f}_c$ are derived from those in (\ref{momentum}) and
(\ref{vsol}). The leading nonlinearities that arise in
(\ref{concsmallfluc}) are of the form
$\nabla_{\perp} \cdot ({\partial_z \bf \delta n}_{\perp} \delta c)$
and similar terms, as well as
$(\nabvec_{\perp} \cdot {\bf \delta n}_{\perp})^2$
and similar terms. These nonlinearities as well as the
couplings to the director in (\ref{concsmallfluc}) are forbidden at
thermal equilibrium.

The director dynamics obtained by inserting (\ref{vsol}) into
(\ref{neom1}) is given to linear order by \beq
\label{dirsmallfluc} {\partial {\bf \delta n}_{\perp} \over \partial t} =
(K_z \partial_z^2 + K_{\perp} \nabla_{\perp}^2 + K^{ \prime}_L 
\nabvec_{\perp} \nabvec_{\perp}
\cdot\,)\, {\bf \delta n}_{\perp} + D_{cn} \partial_z \nabvec_{\perp}
\delta c + {\bf f}_{\perp}
\eeq where, again, the phenomenological parameters and noise source
${\bf f}_{\perp}$ are derived from corresponding terms in
(\ref{momentum}) to (\ref{vsol}). The leading nonlinearities which
will appear in (\ref{dirsmallfluc}) are of the form $\partial_z {\bf
\delta n}_{\perp} (\nabvec_{\perp} \cdot {\bf \delta n}_{\perp})$,
${\bf \delta n}_{\perp} \partial_z ( \nabvec_{\perp} \cdot {\bf \delta
n}_{\perp})$ and similar terms. These nonlinearities too are completely
nonequilibrium in origin.

The concentration field as well as the nonlinearities mentioned above
are absent in the treatment of Gruler et. al. \cite{gruler}. The effect of
the nonlinearities will be dealt with elsewhere \cite{unpub}. The
inclusion of the concentration field leads to our most striking
nonequilibrium effect, namely, the giant number fluctuations we
predict below.

The linearized equations of motion are easily analyzed for their mode
structure, possible instabilities, and fluctuation statistics. This is most
conveniently done by Fourier-transforming in space and time, i.e.,
considering modes that go as $\exp(i {\bf q}.{\bf r} - i
\omega t)$. Doing this enables us to rewrite (\ref{concsmallfluc}) and
(\ref{dirsmallfluc}) as \begin{eqnarray}
\left[-i\omega + D_c \left({\bf \hat{q}} \right) q^2 \right]
\delta c \left({\bf q},\omega \right) + 2 c_0 \alpha q_z\, 
 q_{\perp}\, \delta n_L \left({\bf q}, \omega \right) 
= i{\bf q} \cdot {\bf f}_c\left({\bf q},\omega \right)
\label{c ean}
\end{eqnarray}
\begin{eqnarray} D_{cn}q_z q _{\perp} \delta c \left({\bf q}, \omega
\right) + \left[-i\omega + K_L \left({\bf \hat{q}} \right) q^2
\right] \delta n_{L} \left({\bf q},\omega \right) = f_L
\left({\bf q}, \omega \right)
\label{Long ean}
\end{eqnarray}
\begin{eqnarray}
\left[-i\omega + K_T \left({\bf \hat{q}} \right) q^2 \right]{\bf \delta
n}_T  = {\bf f}_T
\label{Trans ean}
\end{eqnarray} where
\begin{eqnarray} f_L \left({\bf q}, \omega \right) \equiv {\bf
\hat{q}}_{\perp} \cdot {\bf f}_{\perp}\, , \,{\bf f}_T = {\bf f}_{\perp}
- {\bf \hat{q}}_{\perp} \left({\bf \hat{q}}_{\perp}
\cdot {\bf f}_{\perp} \right)
\label{insert 5d1}
\end{eqnarray}
\begin{eqnarray} {\bf \delta n}_L ({\bf q}, \omega) \equiv 
{\bf \hat{q}}_{\perp}\, \left({\bf \hat{q}}_{\perp} 
\cdot {\bf \delta n}_{\perp} ({\bf q}, \omega)\right)
\label{insert 5d1a}
\end{eqnarray} 
and
\begin{eqnarray} {\bf \delta n}_T \equiv {\bf \delta n}_{\perp}
- {\bf \delta n}_{L}
\label{insert 5d1b}
\end{eqnarray} are, respectively, the components of ${\bf f}_{\perp}$
and ${\bf \delta n}_{\perp}$ along and transverse to ${\bf
\hat{q}}_{\perp}$ \cite{d > 2} and we've defined the direction dependent diffusion
constants
\begin{eqnarray} D_c \left({\bf \hat{q}} \right) \equiv D_z
\hat{q}\,^2_z + D_{\perp}
\hat{q}\,^2_{\perp}
\label{Dc}
\end{eqnarray}
\begin{eqnarray} K_L \left({\bf \hat{q}} \right) \equiv K_z
\hat{q}\,^2_z + (K_{\perp}+K^{ \prime}_L)\,
\hat{q} \,^2_{\perp}
\label{KL}
\end{eqnarray}
\begin{eqnarray} K_T \left({\bf \hat{q}} \right) \equiv K_z
\hat{q}\,^2_z + K_{\perp}
\hat{q}\,^2_{\perp}
\label{KT}
\end{eqnarray} The eigenfrequencies obtained from these equations
are all diffusive. There are $d - 2$ transverse diffusive modes coming
from the ${\bf n}_T$ equation, all with identical imaginary
frequencies \begin{eqnarray}
\omega_T = -i K_T \left({\bf \hat{q}} \right) q^2 \label{omega T}
\end{eqnarray} and two coupled concentration - longitudinal modes
with \begin{eqnarray}
\omega_{\pm} = -i \Gamma_{\pm} \left({\bf \hat{q}} \right) q^2
\label{omega pm}
\end{eqnarray} with
\begin{eqnarray}
\Gamma_{\pm} \left({\bf \hat{q}} \right) = {1 \over 2}\left[D_c
\left({\bf \hat{q}} \right) + K_L \left({\bf \hat{q}} \right) \pm
\sqrt{\left(D_c\left({\bf \hat{q}} \right) - K_L\left({\bf \hat{q}}
\right)\right)^2 +8 c_0 \alpha D_{cn}
\hat{q}\, ^2 _{\perp}\hat{q}\, ^2 _z} \,\, \right]
\label{gamma pm}
\end{eqnarray} We shall assume linear stability, which is assured if the
$K_i$s and $D_i$s are positive and the combination $c_0 \alpha D_{cn}$
is not too large.

With these eigenfrequencies in hand, it is straightforward to solve
(\ref{c ean}) -(\ref{Trans ean}) for $\delta c$, $ \delta n_L$, and
${\bf \delta n}_T$ obtaining
\begin{eqnarray}
\delta c = {(2 c_0\alpha q_{\perp}q_z)\, f_L \over \left(\omega + i\Gamma_+
\left({\bf \hat{q}}\right) q^2\right) \left(\omega + i\Gamma_-
\left({\bf \hat{q}} \right) q^2\right)}
\label{delta c prop}
\end{eqnarray}
\begin{eqnarray}
\delta n_L = {-\left(-i \omega + D_c \left({\bf
\hat{q}}\right)q^2\right) f_L \over
\left(\omega + i\Gamma_+
\left({\bf \hat{q}}\right) q^2\right) \left(\omega + i\Gamma_-
\left({\bf \hat{q}}
\right) q^2\right)}
\label{delta nl prop}
\end{eqnarray}
\begin{eqnarray} {\bf \delta n}_T = {{\bf f}_T \over -i\omega + K_T
\left({\bf \hat{q}}\right) q^2}
\label{delta nt prop}
\end{eqnarray} Autocorrelating these, using the known
autocorrelations Eqn.\ (\ref{noise}) for the random forces gives
\begin{eqnarray}
\left< \left|\delta c \left({\bf q}, \omega \right) \right|^2\right> =
{(2 c_0 \alpha q_{\perp} q_z)^2 \,\Delta_n  \over \left(\omega^2 +
\Gamma^2_+ \left({\bf \hat{q}}\right) q^4\right) \left(\omega^2 
+ \Gamma^2_- \left({\bf \hat{q}} \right) q^4\right)}
\label{delta c corr}
\end{eqnarray}
\begin{eqnarray}
\left< \left| \delta n_L \left({\bf q}, \omega \right) \right|^2\right> =
{\left(\omega^2 + D^2_c \left({\bf \hat{q}}\right) q^4 \right)
\Delta_n
\over
\left(\omega^2 +
\Gamma^2_+
\left({\bf \hat{q}}\right) q^4\right) \left(\omega^2 + \Gamma^2_-
\left({\bf \hat{q}}
\right) q^4\right)}
\label{delta nL corr}
\end{eqnarray}
\begin{eqnarray}
\left< \delta n_L \left({\bf q}, \omega \right) \delta c \left(-{\bf q},
-\omega \right) \right> = {2i c_0 \alpha q_z q_{\perp}
\left(\omega + iD_c \left({\bf \hat{q}}\right) q^2\right) \Delta_n
\over \left(\omega^2 + \Gamma^2_+ \left({\bf \hat{q}}\right) q^4\right) 
\left(\omega^2 + \Gamma^2_- \left({\bf \hat{q}} \right) q^4\right)}
\label{delta nL dr corr}
\end{eqnarray}
\begin{eqnarray}
\left< \delta n_{T_{i}} \left({\bf q}, \omega \right) \delta n_{T_{j}}
\left(-{\bf q}, -\omega \right) \right> = {\Delta_n
P^{\perp}_{ij}\left({\bf \hat{q}} \right) \over \left(\omega^2 + K^2_T
\left({\bf \hat{q}}\right) q^4\right)} \label{delta nt corr}
\end{eqnarray} with $\left< \delta n_{T_{i}}\delta n_L\right> =
\langle \delta n_{T_{i}} \delta c \rangle = 0$ for all $i$, ${\bf q}$ and
$\omega$.

In writing eqns.\ (\ref{delta c corr} - \ref{delta nt corr}), we have
neglected all of the contributions coming from the noise ${\bf f}_c$ in
the concentration equation of motion (\ref{vsol}), since these are all
negligible, as ${\bf q}\rightarrow {\bf 0}$, relative to those coming
from the noise ${\bf f}_{\perp}$ in the director equation of motion
(\ref{neom1}).

Equations (\ref{delta c corr} - \ref{delta nt corr}) can now be
Fourier-transformed back to get real space-real time correlations
functions. One quantity of particular interest are the equal-time,
spatially fourier-transformed correlations of the concentration:
\begin{eqnarray}
\left<\delta c ({\bf q}, t) \delta c ({\bf -q}, t) \right> =
\int^{\infty}_{-\infty}{d
\omega
\over 2 \pi}\left<\left| \delta c ({\bf q}, \omega)\right|^2 \right> =
{(2 c_0 \alpha
q_{\perp} q_z)^2 \Delta_n \over F\left({\bf \hat{q}}\right)q^6}
\propto {1 \over q^2}
\label{c equal time}
\end{eqnarray} where
\begin{eqnarray} F\left({\bf \hat{q}}\right) \equiv 2 \left(D_c
\left({\bf \hat{q}} \right) + K_L \left({\bf \hat{q}} \right) \right)
 \left[ D_c \left({\bf \hat{q}} \right)  K_L \left({\bf \hat{q}} \right) 
- 2 c_0 \alpha D_{cn} \hat{q}^2_{\perp} \hat{q}^2_z \right]
\label{hats}
\end{eqnarray}

The important thing about this result is that the $\left(\delta
c\right)^2$  fluctuations are {\it enormous} as $q \rightarrow 0$,
diverging as ${1 \over q^2}$. This would be equivalent, in an
equilibrium system, to having a compressibility which diverged as ${1
\over q^2}$ as $q \rightarrow 0$. To restate this in real space, it is as if
the compressibility of a system of linear extent $L$ diverged according
to $\chi
\propto L^2$ as $L \rightarrow \infty$. Independent of any connection 
to a response quantity like the compressibility,   
the rms number fluctuations $\sqrt{\delta N^2}$ in a volume $V$ 
scale as $\sqrt{S(q \to 0) V}$. For active nematics this implies that 
\begin{eqnarray}
\sqrt{\delta N^2} \propto \sqrt{L^2V} \propto L^{1 + {d \over 2}}
\propto N^{{1 \over 2} + {1 \over d}}
\label{N fluc}
\end{eqnarray} where in the last step we've used the fact that $N
\propto V \propto L^d$ and solved for $L$ as a function of $N$. For
$d = 2$ this recovers our earlier statement that
$\sqrt{\left< \delta N ^2\right>} \propto N$.

Lastly, Eqn.\ (\ref{vsol}) for the velocity implies that the
autocorrelation of a tagged particle will be controlled mainly by the
effect of soft director fluctuations: a particle at a point in the nematic
moves with a speed proportional to the local curvature of ${\bf
n}_{\perp}$. This means the autocorrelation of the velocity ${\bf
v}(t)$ of a particle with trajectory ${\bf R}(t)$ will be, schematically,
\beq
\label{autocorrel}
\langle {\bf v}(0) \cdot {\bf v}(t) \rangle \sim \langle \nabla \delta
n_{\perp} ({\bf R}(0), 0) \nabla \delta n_{\perp}({\bf R}(t), t)\rangle,
\eeq Fourier transforming the ${\bf n}$ correlation functions
(\ref{delta nL dr corr}) and (\ref{delta nt corr}) back to real space and
time, and using the fact that, for a diffusing particle, $\left< \left| {\bf
R}(t) - {\bf R}(0)\right|^2\right> \propto t$, implies that the
correlation function (\ref{autocorrel}) decays as $t^{-d/2}$, as claimed
earlier \cite{diffusionfootnote}.

We close with some comments on the experiments in \cite{kudro}.
Although the nematic ``phase'' in their images appears rather
polydomain, it is likely that their system does display true
nematic order in some parameter range. Perhaps the polydomain
structures coarsen at long times to a true nematic. Such a driven
nematic would be the ideal place to test the predictions of this paper.
If, for some range of parameters, they are able to find macroscopically
aligned regions of {\em tilted} rods --- which would constitute vectorial
order and therefore, as explained in \cite{kudro}, move coherently in
one direction --- these would be examples of the moving XY-model
of \cite{vicseketal,tonertu}. Indeed, the coupled equations
for tilt and concentration in \cite{aranson}, which analyzes the
experiments of \cite{kudro}, are closely related to those of 
\cite{tonertu}. Lastly, the ``vertical rods'' phase, from the 
images, displays a good deal of order; one wonders whether it 
is a nonequilibrium realization of a crystal or hexatic.

\begin{acknowledgments}

SR and JT thank the Aspen Center for Physics, and SR the
Institute for Theoretical Physics, UCSB (NSF Grant PHY99-07949) for partial 
support. JT acknowledges support from NSF grant \#DMR-9980123. 
\end{acknowledgments}



\end{document}